\documentclass[12pt]{article}
\usepackage{epsf}
\usepackage{amsmath}
\usepackage{graphics}
\usepackage{cite}

\renewcommand{\thefootnote}{\#\arabic{footnote}}
\begin{document}

\newcommand{\gtrsim}{ \mathop{}_{\textstyle \sim}^{\textstyle >} }
\newcommand{\lesssim}{ \mathop{}_{\textstyle \sim}^{\textstyle <} }

\newcommand{\rem}[1]{{\bf #1}}

\renewcommand{\thefootnote}{\fnsymbol{footnote}}
\setcounter{footnote}{0}
\begin{titlepage}

\def\thefootnote{\fnsymbol{footnote}}

\hfill arXiv:08mm.nnnn [hep-ph]

\begin{center}

\bigskip
\bigskip
{\Large \bf Simplified Renormalizable $T^{'}$ Model for
Tribimaximal Mixing and Cabibbo Angle}

\vspace{1.0in}

{\bf Paul H. Frampton$^{(a)}$\footnote{frampton@physics.unc.edu}
, Thomas W. Kephart$^{(b)}$\footnote{tom.kephart@gmail.com}
and Shinya Matsuzaki$^{(a)}$\footnote{synya@physics.unc.edu}}

\vspace{1.0in}

{$^{(a)}$ Department of Physics and Astronomy, UNC-Chapel Hill, NC 27599.}

{$^{(b)}$ Department of Physics and Astronomy, Vanderbilt University,
Nashville, TN 37235.}

\end{center}

\vspace{1.0in}

\begin{abstract}
In a simplified renormalizable model where
the neutrinos have PMNS (Pontecorvo-Maki-Nakagawa-Sakata)
mixings tan$^{2} \theta_{12} = \frac{1}{2}, \theta_{13}=0,
\theta_{23} = \pi/4$ and with flavor symmetry
$T^{'}$
there is a corresponding prediction
where the quarks have CKM (Cabibbo-Kobayashi-Maskawa)
mixings tan$2 \Theta_{12} = \frac{\sqrt{2}}{3}, \Theta_{13}=0,
\Theta_{23} =0$.
\end{abstract}

\end{titlepage}

\renewcommand{\thepage}{\arabic{page}}
\setcounter{page}{1}
\renewcommand{\thefootnote}{\#\arabic{footnote}}

The standard model of particle physics is very predictive, and 
well tested. Given coupling constants and masses, we can calculate 
electroweak processes like scattering cross-sections and atomic 
energy levels to remarkable accuracy. We can in principle also 
calculate hadronic processes except for our lack of technical 
skill, but not for our lack of a good theory. The main reason 
for the precision (or potential precision) in these calculations 
is the symmetries satisfied by the theory--Lorentz invariance 
plus the electroweak gauge symmetry $SU(2)\times U(1)$ and 
the color sector gauge symmetry $SU(3)$ in addition to the 
discrete spacetime symmetries like $P$ and $CP$, and the 
controlled way some of these symmetries are broken, e. g., 
gauge symmetry breaking via Higgs vacuum expectation value 
in the electroweak sector. However, the standard model has 
its limitations. There are still approximately twenty parameters 
needed as input for the model, including gauge coupling constants, 
quark and lepton mixing angles and phases, etc.
 
Recently, a considerable effort has been made to reduce the 
number of standard model input parameters. This can be done by 
introducing new symmetries that relate the various parameters, 
while eventually breaking or at least partially breaking the 
new symmetry. Some requirements and constraints needed for a 
viable theory are: the new symmetries cannot be gauged at 
low energy since there are no corresponding light gauge 
bosons in the spectrum. Broken continuous global symmetries 
must also be avoided since they lead to Goldstone bosons, 
also unseen in experiments. This leads us to the only 
natural choice--discrete symmetries.
 
To date, models of this type have usually focused on 
reducing the number of parameters in either the lepton or the quark sector. 
A notable exception is provided by models based on the binary tetrahedral 
group $T^{'}$, which is capable of providing calculability to both 
sectors. To show the power of this additional symmetry, we will 
provide a $T^{'}$ model that leads to the celebrated tribimaximal 
neutrino mixing and at the same time allows us to calculate quark 
mixings. As an example, we will show how the quark mixing matrix 
can give a purely numerical value for the Cabibbo angle that is 
only a few percent away from its experimental value.

\bigskip

The first use of the binary tetrahedral
group $T^{'}$ in particle physics was by
Case, Karplus and Yang\cite{Yang}
who were motivated to consider gauging a finite
$T^{'}$ subgroup of $SU(2)$ 
in Yang-Mills
theory. 
This led Fairbairn, Fulton and Klink (FFK) \cite{FFK}
to make an analysis
of $T^{'}$ Clebsch-Gordan coefficients
\footnote{Other analysis of $T^{'}$ Clebsch-Gordan
coefficients appears in \cite{Y,ACL,Feruglio}. We 
use FFK.}.
As a flavor symmetry, $T^{'}$ first
appeared in \cite{FK1} motivated by 
the idea of representing the three quark families
with the third treated differently from
the first two. 
Since $T^{'}$ is the double cover
of $A_4$, it was natural to suggest\cite{FK2} that
$T^{'}$ be employed to accommodate quarks
and simultaneously 
the established $A_4$ model building
for tribimaximal neutrino mixing.

\bigskip

In the present article we shall build such a $T^{'}$ model 
with simplifications to emphasize the largest quark mixing, the
Cabibbo angle,
for which we shall derive an entirely new formula as an exact angle.  

\bigskip

This work is a 
major extension of that in \cite{FM}
where the constraint of renormalizability was first applied 
to an $A_4$ model and led not only to the usual
tribimaximal mixing\footnote{Throughout we ignore CP violation.}
\begin{equation}
\tan\theta_{12} = 1/\sqrt{2}, ~~~~ \theta_{23} = \pi/4,
~~~~~ \theta_{13}=0,
\label{tribimaximal}
\end{equation}
but to
the 
simplified normal
hierarchy 
\begin{equation}
m_3 \neq 0,  ~~~~~ m_{1,2} =0.
\label{normalhierarchy}
\end{equation}

\noindent We review briefly this $A_4$ model. 
The leptons are assigned 
under ($A_4 \times Z_2$) as

\begin{equation}
\begin{array}{ccc}
\left. \begin{array}{c}
\left( \begin{array}{c} \nu_{\tau} \\ \tau^- \end{array} \right)_{L} \\
\left( \begin{array}{c} \nu_{\mu} \\ \mu^- \end{array} \right)_{L} \\
\left( \begin{array}{c} \nu_e \\ e^- \end{array} \right)_{L} 
\end{array} \right\} 
L_L  (3, +1)  &
\begin{array}{c}
~ \tau^-_{R}~ (1_1, -1)   \\
~ \mu^-_{R} ~ (1_2, -1) \\
~ e^-_{R} ~ (1_3, -1)  \end{array}
&
\begin{array}{c}
~ N^{(1)}_{R} ~ (1_1, +1) \\
~ N^{(2)}_R ~ (1_2, +1) \\
~ N^{(3)}_{R} ~ (1_3, +1),\\  \end{array}
\end{array}
\end{equation}
which is typical of $A_4$ model building\cite{Ma}. 
Imposing strict renormalizability on the lepton lagrangian 
allows as nontrivial terms 
only Majorana mass terms and Yukawa couplings 
to $A_4$ scalars
\footnote{All scalars are doublets under electroweak $SU(2)$.}
$H_3(3,+1)$ and $H_3^{'}(3,-1)$ 

\begin{eqnarray}
{\cal L}_Y^{(leptons)}
&=&
\frac{1}{2} M_1 N_R^{(1)} N_R^{(1)} + M_{23} N_R^{(2)} N_R^{(3)} \nonumber \\
& & + \Bigg\{
Y_{1} \left( L_L N_R^{(1)} H_3 \right) + Y_{2} \left(  L_L N_R^{(2)}  H_3
\right) + Y_{3}
\left( L_L N_R^{(3)} H_3 \right)  \nonumber \\
&& +
Y_\tau \left( L_L \tau_R H'_3 \right)
+ Y_\mu  \left( L_L \mu_R  H'_3 \right) +
Y_e \left( L_L e_R H'_3 \right)
\Bigg\}
+
{\rm h.c.}.
\label{Ylepton}
\end{eqnarray}

\noindent Charged lepton masses arise from the vacuum expectation value 
(hereafter VEV)

\begin{equation}
<H_3^{'}> = 
\left(\frac{m_{\tau}}{Y_{\tau}},\frac{m_{\mu}}{Y_{\mu}},\frac{m_{e}}{Y_{e}}
\right) = ( M_{\tau}, M_{\mu}, M_e ),
\label{Hprime}
\end{equation}
where $M_i \equiv m_i/Y_i$ ($i = \tau, \mu, e$).
Neutrino masses and mixings satisfying
Eqs.(\ref{tribimaximal},\ref{normalhierarchy}) 
come from the see-saw mechanism\cite{Minkowski}
and the VEV 
\footnote{Use \cite{TDLee} of $<H_3>=V(1, 1, 1)$ gives 
Eqs.(\ref{tribimaximal},\ref{normalhierarchy}) 
with $2\leftrightarrow3$ interchanged in Eq.(\ref{normalhierarchy}).}
\begin{equation}
<H_3> = V( 1, -2, 1).
\label{VEV}
\end{equation}

\noindent We promote $A_4$ to $T^{'}$ keeping renormalizability and including quarks.
The left-handed quark doublets \noindent $(t, b)_L, (c, d)_L, (u, d)_L$
are assigned under $(T^{'} \times Z_2)$ to

\begin{equation}
\begin{array}{cc}
\left( \begin{array}{c} t \\ b \end{array} \right)_{L}
~ {\cal Q}_L ~~~~~~~~~~~ ({\bf 1_1}, +1)   \\
\left. \begin{array}{c} \left( \begin{array}{c} c \\ s \end{array} \right)_{L}
\\
\left( \begin{array}{c} u \\ d \end{array} \right)_{L}  \end{array} \right\}
Q_L ~~~~~~~~ ({\bf 2_1}, +1),
\end{array}
\label{qL}
\end{equation}

\noindent and the six right-handed quarks as

\begin{equation}
\begin{array}{c}
t_{R} ~~~~~~~~~~~~~~ ({\bf 1_1}, +1)   \\
b_{R} ~~~~~~~~~~~~~~ ({\bf 1_2}, -1)  \\
\left. \begin{array}{c} c_{R} \\ u_{R} \end{array} \right\}
{\cal C}_R ~~~~~~~~ ({\bf 2_3}, -1)\\
\left. \begin{array}{c} s_{R} \\ d_{R} \end{array} \right\}
{\cal S}_R ~~~~~~~~ ({\bf 2_2}, +1).
\end{array}
\label{qR}
\end{equation}

\noindent We add only two new scalars $H_{1_1} (1_1, +1)$ and
$H_{1_3} (1_3, -1)$ whose VEVs
\begin{equation}
<H_{1_1}> = m_t/Y_t, ~~~~ <H_{1_3}> = m_b/Y_b,
\label{H13VEV}
\end{equation}
provide the $(t, b)$ masses.
In particular, no  $T^{'}$ doublet
($2_1, 2_2, 2_3$) scalars have been added. This 
allows a non-zero value only for $\Theta_{12}$. The other angles  
vanish making the third family stable
\footnote{At the end of this paper non-vanishing $\Theta_{23}, \Theta_{13}$ is related to $(d, s)$ masses.}.
The allowed quark Yukawa and mass terms are

\begin{eqnarray}
{\cal L}_Y^{(quarks)}
&=& Y_t ( \{{\cal Q}_L\}_{\bf 1_1}  \{t_R\}_{\bf 1_1} H_{\bf 1_1}) \nonumber \\
&&
+ Y_b (\{{\cal Q}_L\}_{\bf 1_1} \{b_R\}_{\bf 1_2} H_{\bf 1_3} ) \nonumber \\
&& 
+ Y_{{\cal C}} ( \{ Q_L \}_{\bf 2_1} \{ {\cal C}_R \}_{\bf 2_3} H^{'}_{\bf 3})
\nonumber \\
&& 
+ Y_{{\cal S}} ( \{ Q_L \}_{\bf 2_1} \{ {\cal S}_R \}_{\bf 2_2} H_{\bf 3})
\nonumber \\
&&
+ {\rm h.c.}.
\label{Yquark} 
\end{eqnarray} 

The use of $T^{'}$ singlets and doublets
\footnote{It is discrete anomaly free cf. \cite{Ramond,Luhn}.
We thank the UF-Gainesville group for discussions.}
for quark families 
in Eqs.(\ref{qL},\ref{qR})
permits the third family to differ 
from the first two and thus make plausible the 
mass hierarchies $m_t \gg m_b$,  $m_b > m_{c,u}$ 
and $m_b > m_{s,d}$ as outlined in \cite{FK1}. 

The nontrivial ($2 \times 2$) quark mass matrices 
$(c, u)$ and $(s, d)$ will be respectively denoted by $U^{'}$ 
and $D^{'}$ and calculated using the $T^{'}$ Clebsch-Gordan
coefficients of FFK\cite{FFK}. Dividing out $Y_{{\cal C}}$ and $Y_{{\cal S}}$ 
in  Eq.(\ref{Yquark}) gives $U$ and $D$ 
matrices ($\omega = e^{i \pi/3}$)

\begin{equation}
U \equiv \left( \frac{1}{Y_{{\cal C}}}\right) U^{'}  = \left(\begin{array}{cc}
\sqrt{\frac{2}{3}} \omega^2 M_{\tau} & \frac{1}{\sqrt{3}} M_e \\
- \frac{1}{\sqrt{3}} \omega^2 M_e & \sqrt{\frac{2}{3}} M_{\mu} 
\end{array}
\right),
\label{Umatrix}
\end{equation}

\begin{equation}
D \equiv \left( \frac{1}{V Y_{{\cal S}}} \right)
D^{'}  = \left( \begin{array}{cc}
\frac{1}{\sqrt{3}} & - 2 \sqrt{\frac{2}{3}} \omega \\
\sqrt{\frac{2}{3}} & \frac{1}{\sqrt{3}} \omega 
\end{array}
\right).
\label{Dmatrix}
\end{equation}

\noindent Let us first consider $U$ of Eq.(\ref{Umatrix}).
Noting that $m_{\tau} > m_{\mu} \gg m_e$
we may simplify $U$ by setting the electron mass
to zero, $M_e = 0$. This renders $U$ diagonal leaving
free the c, u, $\tau$ and $\mu$ masses.
This leaves only the matrix $D$ in Eq.(\ref{Dmatrix})
which predicts both $\Theta_{12}$ and $(m_d^2/m_s^2)$.
The hermitian square 
${\bf {\cal D}} \equiv D D^{\dagger}$
is 
\begin{equation}
{\bf {\cal D}} \equiv D D^{\dagger} = \left(
\frac{1}{3} \right) \left(
\begin{array}{cc}  9 & - \sqrt{2}  \\
- \sqrt{2} & 3 
\end{array}
\right),
\label{DDdagger}
\end{equation}
which leads by diagonalization to a formula for the Cabibbo angle

\begin{equation}
\tan 2\Theta_{12} = \left( \frac{\sqrt{2}}{3} \right),
\label{Cabibbo}
\end{equation}

\noindent or equivalently\footnote{Ellipsis ... denotes exactitude.}
$\sin \Theta_{12} = 0.218...$
close to the experimental value
\footnote{Experimental results are from \cite{PDG2006};
see references therein.}
$\sin \Theta_{12} \simeq 0.227$.

\bigskip

Our result of an exact angle for $\Theta_{12}$ can be regarded
as on a footing
with the tribimaximal values for neutrino angles $\theta_{ij}$,
quoted in Eq.(\ref{tribimaximal}).
Note that the tribimaximal $\theta_{12}$ 
presently agrees with experiment within one standard
deviation ($1 \sigma$).  On the other hand,
our analagous exact angle for $\Theta_{12}$
differs from experiment already by $9 \sigma$ which is
probably a reflection of the fact
that the experimental accuracy for $\Theta_{12}$ is $\sim 0.5\%$
while that for $\theta_{12}$ is $\sim 6\%$. 
It is thus very important 
to acquire better experimental data on $\theta_{12}$,
$\theta_{23}$ and $\theta_{13}$
to detect their similar deviation from the exact angles
predicted by Eq.(\ref{tribimaximal}).
Our result for $(m_d^2/m_s^2)$ from Eq.(\ref{DDdagger}) is exactly
$0.288...$ compared to the central experimental value
$\simeq 0.003$ in a simplified
model whose generalization to an extended 
scalar sector including $T^{'}$ doublets can 
avoid $\Theta_{23} = \Theta_{13} = 0$ 
and thereby change $(m_d^2/m_s^2)$ 
due to mixing of $(d, s)$
with the $b$ quark.

\bigskip
 
We believe our $T'\times Z_2$ extension of the standard model is an 
important stride in tying the quark and lepton sectors together, 
providing calculability, and at the same time reducing the number 
of standard model parameters. The ultimate goal would be to 
understand the origin of this discrete symmetry. Since gauge 
symmetries can break to discrete symmetries, and gauge symmetries 
arise naturally from strings, perhaps there is a clever 
construction of our model with its fundamental origin in string theory.

\newpage

\begin{center}
{\bf Acknowledgements}
\end{center}

\noindent This work was supported by U.S. Department of Energy grants number 
DE-FG02-05ER41418 (P.H.F. and S.M) and  DE-FG05-85ER40226 (T.W.K.).

\end{document}